\documentclass[prl,twocolumn,showpacs,preprintnumbers,amsmath,amssymb]{revtex4}

\usepackage{graphicx}
\usepackage{dcolumn}
\usepackage{bm}
\usepackage[]{ams,amsmath,amssymb,epsfig}

\begin{document}
\title{Light-induced  valley currents and magnetization in graphene rings}

\author{A.S. Moskalenko}
\email{andrey.moskalenko@physik.uni-halle.de}
\altaffiliation[Also
at ]{A.F. Ioffe Physico-Technical Institute, 194021 St.
Petersburg, Russia}
\author{J. Berakdar}
\affiliation{Institut f\"ur Physik,
Martin-Luther-Universit\"at
 Halle-Wittenberg,  Nanotechnikum-Weinberg, Heinrich-Damerow-St. 4, 06120 Halle,
 Germany}

\date{\today}

\begin{abstract}
We study the non-equilibrium  dynamics  in a mesoscopic graphene ring excited by picoseconds shaped electromagnetic pulses. We predict an ultrafast buildup of
 charge polarization, currents and orbital magnetization.  Applying the  light pulses
  identified here, non-equilibrium valley currents are  generated in a graphene ring threaded by a stationary magnetic flux. We predict a finite  graphene ring  magnetization even for a
   vanishing charge current; the magnetization emerges due to the light-induced difference of the valley populations.
\end{abstract}

\pacs{73.63.-b,73.23.-b,73.22.Gk,81.05.Uw}

\maketitle

\textit{Introduction.-}
  Since the recent  fabrication of graphene,  a
  monolayer of carbon, a
  number of  fascinating phenomena have been uncovered, mostly owing to
  the quasi-relativistic behavior of the carriers and their high mobility \cite{Castro_Neto2009,Novoselov_Nature2005,Geim2007,Beenakker_RMP2008,Morozov2008}.
  Two $\sigma$-bonded interpenetrating triangular sublattices, $A$ and $B$, build the graphene honeycomb lattice.
  The $\pi$ and $\pi^*$  bands govern the electronic properties near the neutrality point and result in
  conical valleys touching  at the high symmetry points $K$ and $K'$ of the Brillouin zone (BZ).
   Near $K$ and $K'$
   the energy dispersion is linear and the electronic properties are well described
by the effective Dirac-Weyl Hamiltonian $H_0$ \cite{Novoselov_Nature2005,Geim2007,Beenakker_RMP2008}
\footnote{$H_0=v\vec{\sigma}\cdot\vec{p}$,
where $\vec{p}$ is the momentum operator,  $\vec{\sigma}=(\sigma_x,\sigma_y)$ and $v$ is the Fermi velocity.  $\sigma_x$ and $\sigma_y$ are Pauli matrices  built on the basis of the pseudospin wavefunctions corresponding to the different sublattices.\label{note}}.
The  stationary  states are  degenerate in the spin ${\cal S}$ and the valley  quantum numbers $\tau=\pm 1$. The latter correspond to the two non-equivalent $K$-points in BZ.
Due to the suppressed intervalley scattering  the control of $\tau$ eigenstates  may be utilized for novel electronic \cite{Rycerz2007,Xiao2007} and optoelectronic applications \cite{Wang2008}.
  New physical effects emerge due to confinement. E.g.,  in a  mesoscopic graphene rings
pierced by a magnetic flux, the ring confinement breaks the valley degeneracy and  results
 in the persistent current  \cite{Recher2007}. Experimentally, such
 graphene rings were fabricated and the Aharonov-Bohm effect was observed \cite{Russo2008}.

 While a large body of work has been devoted to various  equilibrium electronic  and optical properties,
 the non-equilibrium time-dependent phenomena in graphene are much less explored \cite{Mikhailov2008,Syzranov2008}.
 The present paper presents the first study  on the
  non-equilibrium dynamics in graphene rings driven by
asymmetric monocycle electromagnetic pulses.
 Charge polarization and current carrying states build up within picoseconds and
 are tunable by the parameters of the driving field. The states may become valley polarized resulting in a non-equilibrium valley currents. The valley population together with the charge current determine the magnetization of the ring.
\textit{Stationary states.-}
We consider a graphene ring \cite{Recher2007} of radius $r_0$ and width $W$ (cf. Fig.\ref{fig2}(a)) threaded by a magnetic flux of a strength $\Phi$.  As in \cite{Berry1987,Recher2007,Akhmerov2008}, the
Dirac electrons are confined to the ring by the potential $\tau V(r)\sigma_z$  at
 the boundaries, as resulting e.g. from  a substrate potential \cite{Zhou2008,Rotenberg_2008}.
  The polar-coordinates ring hamiltonian is    [29]
\begin{equation}\label{Eq:Hamiltonian_mod}
    H=-i\hbar v\left[\sigma_r\partial_r+\sigma_\phi\frac{1}{r}(\partial_\phi+i\tilde{\Phi})\right]+\tau V(r)\sigma_z\;,
\end{equation}
where $\sigma_r=\vec{\sigma}\cdot \vec{e}_r$ and $\sigma_\phi=\vec{\sigma}\cdot \vec{e}_\phi$ with $\vec{e}_r$ and $\vec{e}_\phi$ being the basis vectors of the polar coordinate system and
$\sigma_z$ is the Pauli matrix expressed in the pseudospin states  of the two sublattices. $v=10^6$~m/s is the Fermi velocity and
 $\tilde{\Phi}=\Phi/\Phi_0$,  where  $\Phi_0$ 
is the flux quantum.
The eigenstates of $H$ and  $J_z$ (the $z$ component of
the total angular momentum with eigenvalues $m$) are
\begin{equation}\label{Eq:psi_phi_sep}
    \psi_m(r,\phi)=R_+(r) e^{i(m-1/2)\phi}\chi_+ +R_-(r) e^{i(m+1/2)\phi}\chi_-\;,
\end{equation}
where $m=\pm\frac{1}{2},\pm\frac{3}{2},...$  and 
$\sigma_z\chi_{\pm}=\pm\frac{1}{2}\chi_{\pm}$.
The general form for the radial parts is $R_+(r)=c_1H^{(1)}_{\overline{m}-\frac{1}{2}}(\tilde{r})+c_2H^{(2)}_{\overline{m}-\frac{1}{2}}(\tilde{r})$ and $R_-(r)=i s \left[c_1 H^{(1)}_{\overline{m}+\frac{1}{2}}(\tilde{r})+c_2H^{(2)}_{\overline{m}+\frac{1}{2}}(\tilde{r})\right]$,
where $\tilde{r}=r |E|/v$ is a normalized radial coordinate, $H_m^{(1)[(2)]}$ is the Hankel function of the first [second] kind,  $s=\mbox{sgn}(E)$ selects  the solution of the positive or  the negative energy branch, and $\overline{m}=m+\tilde{\Phi}$.
The boundary conditions and the normalization fix the coefficients $c_1$ and $c_2$.
For $V(r)=0$ if $r\in\left(r_0-\frac{W}{2},r_0+\frac{W}{2}\right)$, and $V(r)\rightarrow +\infty$ outside the ring \cite{Recher2007} we find $\psi=\pm\tau\sigma_\phi\psi$ at $r=r_0\pm\frac{W}{2}$.
With this  Eq.~\eqref{Eq:psi_phi_sep} can be solved numerically or for $W/r_0\ll 1$ analytically \cite{Recher2007} yielding
the spectrum
\begin{eqnarray}\label{Eq:Energy_limit_rewritten}
    E_{nm}^{s\tau}&=&s\varepsilon_n+s\lambda_n\left[m+\tilde{\Phi}^{s\tau n}_{\rm eff}\right]^2-\frac{s\lambda_n}{4\pi^2(n+1/2)^2},\\
´
 \varepsilon_n&=&\frac{\hbar v}{W}(n+1/2), \lambda_n=\frac{\hbar v}{W}\left(\frac{W}{r_0}\right)^2\frac{1}{\pi(2n+1)}.\label{Eq:Energy_limit_rewritten2}\end{eqnarray}
  $n=0,1,2,...,$ and
$\tilde{\Phi}^{s\tau n}_{\rm eff}=\tilde{\Phi}-\frac{1}{2}\frac{s\tau}{(n+1/2)\pi}$.
For fixed $s,\tau$, and $n$ the quantity $\tilde{\Phi}^{s\tau n}_{\rm eff}$ modifies the energy spectrum as an effective (normalized) magnetic flux. The shift of the effective magnetic flux from $\tilde{\Phi}$ has a different sign depending on  the valley ($\tau=\pm 1$).
For $W/r_0\ll 1$ we find
\begin{eqnarray}
    R_{+,n}^{s\tau}(r)=\frac{1}{\sqrt{Wr_0}}\cos\left[(n\!+\!1/2)\pi\tilde{r}'\!-\!\frac{\tau}{4}\pi\right]\;,
    \label{Eq:R+limit}\\
    R_{-,n}^{s\tau}(r)=\frac{is}{\sqrt{Wr_0}}\sin\left[(n\!+\!1/2)\pi\tilde{r}'\!-\!\frac{\tau}{4}\pi\right]\;,
    \label{Eq:R-limit}
\end{eqnarray}
where $\tilde{r}'=\left(r-r_0+\frac{W}{2}\right)/W\in(0,1)$.
For applications involving tunneling from the ring it is important to inspect the case of
a  finite barrier boundary, i.e. $V=V_0$ for $r$ outside of the ring.  To a
 first order of $\gamma\equiv \frac{\hbar v}{WV}\ll 1$ we find that  Eq.~\eqref{Eq:Energy_limit_rewritten2} applies with $\varepsilon_n$ being replaced  by $(1-\gamma)\varepsilon_n$ and
$\lambda_n$ by $(1+\gamma)\lambda_n$. For $W=0.1$ $\mu$m the
condition $\gamma\ll 1$ means $V\gg 7$~meV. Hence, our theory
developed below is valid also for a finite barrier graphene ring.
In a particular example of the boron nitride substrate we have
$V=53$~meV \cite{boron_nitride} and therefore $\gamma=0.13$.
%
Having specified the stationary single-particle states
we proceed with the non-equilibrium calculations
\footnote{For few-electron rings the Coulomb interaction may  influence the  equilibrium properties \cite{Abergel2008} but  for a larger particle numbers it is less significant.}.

\begin{figure}
  \includegraphics[width=6.0cm]{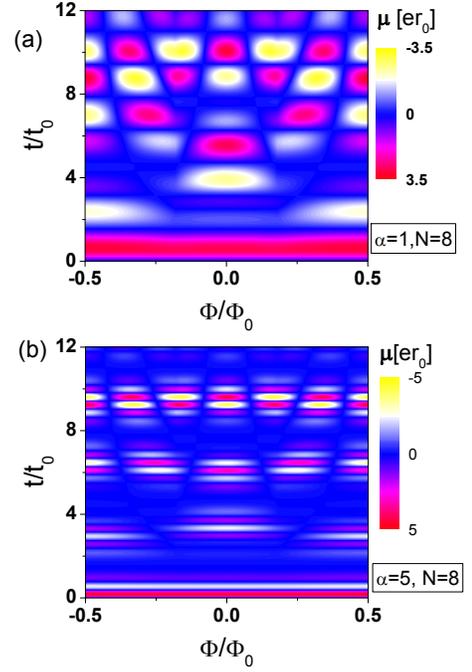}\\
  \caption{Dependence of the dipole moment generated in the graphene ring on the time past after the excitation and the normalized magnetic flux $\tilde{\Phi}=\Phi/\Phi_0\in[-1/2,1/2]$ (outside of this range the periodicity by $\tilde{\Phi}\rightarrow \tilde{\Phi}+1$ can be used)  in the case of  $8$ carriers in the ring at $T=0$ for (a) $\alpha=1$ and (b) $\alpha=5$.}\label{mu_phi_t}
\end{figure}

\begin{figure}
  \includegraphics[width=8.5cm]{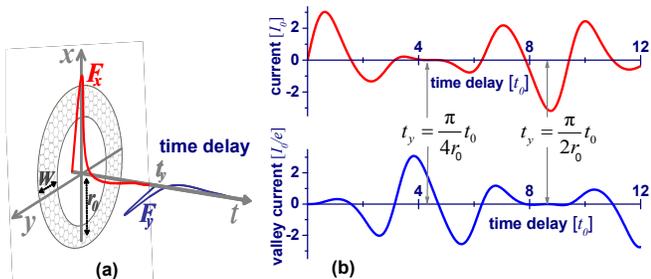}\\
  \caption{(a) Current generation in the ring by application of two HCPs polarized along mutually perpendicular directions and delayed in respect to each other by the time $t_y$. (b) Upper and lower figures show the total current and the valley current, respectively, generated in the graphene ring in dependence on the delay time $t_y$. Value of the magnetic flux is set to $\Phi=\Phi_0/\pi$, number of carriers is $N=8$, excitation strengths of both HCPs are equal to $\alpha=1$. Arrows   at $t_y=\frac{\pi}{4r_0}t_0$ indicates the delay time for which a valley current with no total charge current is generated whereas for a delay time $t_y=\frac{\pi}{2r_0}t_0$ a total charge current with equal contributions from both valleys is generated.}\label{fig2}
\end{figure}

\textit{ Pulse-induced polarization.-}
To drive the non-equilibrium states in graphene rings we utilize
asymmetric monocycle  pulses, so-called  half-cycle pulses (HCPs) \cite{You1993,Jones1993}.
For pulse duration $\tau_p$  shorter than the carriers characteristic time scale \footnote{
For  rings with $W/r_0\ll 1$ IA requires $\tau_p\ll \hbar/\lambda_n$ for all radial channels influenced by the excitation.}
 the impulsive approximation
 (IA) applies,
meaning that
 the time-dependent  carrier wavefunction $\Psi(r,\varphi;t)$ propagates  stroboscopically as ~\cite{Alex_PRB2004}
\begin{equation}
\begin{split}
& \Psi(r,\varphi,t^{+})=\Psi(r,\varphi,t^{-})\ e^{i\vec{\alpha}\vec{e}_r};\,\\
& \vec{\alpha}=\frac{r_0 \vec{p}}{\hbar}\;,\ \ \ \ \ \ \ \vec{p}=e\int \vec{F}(t)dt\;,
\end{split}
\label{eq:matching}
\end{equation}
where  $t^{-}$ and $t^{+}$ refer to times  before and after the application of the  pulse and
$\vec{\alpha}$ is the action  delivered to a ring carrier of charge $e$  by
the HCP  electric field  $F(t)$.
The  pulse triggers  a   time-dependent carrier density distribution which depends on $\tau=\pm 1$, i.e. it is different for the two valleys.
As a physical consequence,
a time-dependent charge dipole moment is created in the ring. For the post-pulse dipole moment $\mu^{\tau_0}_{m_0}(t)$
associated with a carrier starting from the stationary state  $\tau=\tau_0$ and $m=m_0$
we find \footnote{We limit
the considerations to the lowest radial channel of the positive energy branch, i.e.
$n=0$ and $s=1$. This is achieved experimentally by applying  a gate voltage.}
\begin{equation}\label{Eq:mu_tau_m0}
    \mu^{\tau_0}_{m_0}(t)=er_0\alpha h(\Omega)\sin\frac{t}{t_0}
    \cos\left[2(m_0+\Phi^{\tau_0}_{\rm eff})\frac{t}{t_0}\right],
\end{equation}
where  $t_0=\hbar/\lambda_0$, $h(\Omega)=J_0(\Omega)+J_2(\Omega)$, $J_l(x)$ denotes the
Bessel function of the order $l$,
and $\Omega=\alpha\sqrt{2\left[1-\cos (2t/t_0)\right]}$.
The total electric dipole created in the ring for a fixed $\tau$, spin value ${\cal S}$, and $t>0$ is  $\mu^{\tau}(t)=\sum_m f^\tau_{m} \mu^{\tau}_{m}(t)$, where $f^\tau_{m}$ is the equilibrium distribution function.
For $N_\tau$ carriers in a given valley $\tau$ at zero temperature $T=0$ we carried out the summation over $m$ analytically.
 For an arbitrary even or odd $N_\tau$  we find respectively
\begin{eqnarray}\label{Eq:mu_tau_T0_even}
  \!\!\! &&\!\!\! \mu^{\tau}_{\rm even}(t)\!=er_0\alpha h(\Omega)
    \cos\!\left[\frac{2t}{t_0}\left(\left|\tilde{\Phi}^\tau_{\rm eff}\right|-\frac{1}{2}\right)\right]
    \sin\frac{N_\tau t}{t_0},\nonumber\\
&&\!\!\!\!\mu^{\tau}_{\rm odd}(t)=er_0\alpha h(\Omega)
    \cos\!\left[\frac{2t}{t_0}\tilde{\Phi}^\tau_{\rm eff}\right]
    \sin\frac{N_\tau t}{t_0}.
\label{Eq:mu_tau_T0_odd}\end{eqnarray}
 Both expressions apply for $\tilde{\Phi}^\tau_{\rm eff}=\tilde{\Phi}-\frac{\tau}{\pi}\in[-1/2,1/2]$, outside of this interval $\mu^{s\tau}(t)$ is determined from the periodicity
  in $\tilde{\Phi}^\tau_{\rm eff}$ with a period 1.
The total dipole moment  $\mu(t)$ depends on the distribution of the carriers between the valleys that in turn depends on the magnetic flux $\tilde{\Phi}$.
 The spin degeneracy  is also important. One can show that  jumps in the population of particular states take place only at the points $\tilde{\Phi}=-\frac{1}{2},-\frac{1}{\pi},-r,0,r,\frac{1}{\pi},\frac{1}{2}$
for $\tilde{\Phi}$ in the interval $[-1/2,1/2]$, where we denote $r=\frac{1}{2}-\frac{1}{\pi}$.
The dynamics of the dipole moment for $N=8$ carriers is shown in Fig.~\ref{mu_phi_t} as a function of the applied stationary  magnetic flux for two different excitation strengths $\alpha=1$ and $\alpha=5$, showing that the
ring electric dipole  and hence the associated light emission are controllable by $\Phi$ and $\alpha$.
For $r_0=1~\mu$m and HCPs with a sine-square
shape and a time duration of 0.5~ps, $\alpha=1$ corresponds to the peak value of electric field $F=26$~V/cm.\\
Note,   the boundary conditions break the effective time-reversal symmetry \cite{Recher2007} making
the states corresponding to different $\tau$   but otherwise to the same quantum numbers non-degenerate.  The dynamics of the charge polarization for confined carriers is however the same in both $\tau$ valleys for $\Phi=0$. This  follows from the invariance  of the states under $\tau\rightarrow -\tau$, and  $m \rightarrow -m$ at $\tilde{\Phi}=0$, as evidenced by  Eqs.~\eqref{Eq:mu_tau_T0_even}. This degeneracy is lifted by applying a stationary magnetic flux $\tilde{\Phi}\neq 0$. The density distribution of carriers in the ring becomes valley-polarized.\\
\textit{Non-equilibrium charge and valley currents.-}
The electric current density
  $\vec{j}=ev\psi^\dagger \vec{\sigma} \psi$ has
the $\phi$-component of the current density
\begin{equation}\label{Eq:current_density}
   j_\phi(r)=-2{\rm Im}[R_+(r)R_-^*(r)]\;.
\end{equation}
The total charge current is  $I=\int j_\phi{\rm d}r$.
To a zero order in $W/r_0\ll 1$ we find
 $I=0$. Only higher order corrections in $W/r_0$ give rise to  a non-vanishing ring current. For the eigenstate specified by $s,\tau,n,m$ the lowest order correction \cite{Recher2007} to the current follows   from $I^{s\tau}_{nm}=-\partial E_{nm}^{s\tau}/\partial \Phi $ using the energy spectrum in the considered limit \footnote{We  checked numerically   the validity of this  formula for $I^{s\tau}_{nm}$ in the limit case $W/a\ll 1$ by comparing with the general equation \eqref{Eq:current_density}.}.
For  $n=0$, $s=1$ this current is equal to $I^{\tau}_{m}=-I_0\left(m+\tilde{\Phi}^\tau_{\rm eff}\right)$, where $I_0=\frac{1}{\pi^2}\frac{|e|vW}{r_0^2}$.
For a ring with  $r_0=1~\mu\mbox{m}$ and $W=100~\mbox{nm}$ we obtain $I_0=0.16~\mbox{nA}$, if $r_0=425~\mbox{nm}$ and $W=150~\mbox{nm}$ as in Ref.~\onlinecite{Russo2008} we find $I_0=1~\mbox{nA}$. In both cases IA is valid if $\tau_p \ll 3~\mbox{ps}$. Such HCPs are experimentally available \cite{You1993}.
The total current in the ring $I$ is the sum of an  equilibrium (persistent) current $I_{\rm eq}$
 and a non-equilibrium time-dependent current part $I_{\rm neq}(t)$ generated in the ring: $I=I_{\rm eq}+I_{\rm neq}(t)$. The equilibrium part is given by $I_{\rm eq}=\sum_{m\tau} f^\tau_{m\tau} I^{\tau}_{m}$. For $T=0$ it is given  in Ref.~\onlinecite{Recher2007} for $N=1,2,3,4$. We derive it for any  $N$ in $n=0$.

A non-equilibrium ring current is generated by  a sequence of two time-delayed mutually perpendicular HCPs (see Fig.~\ref{fig2}(a)), similarly to the pulse-current generation in semiconductor rings \cite{Alex_PRL2005,Alex_Europhysics2005,Moskalenko_PRB2006,Moskalenko_PRA2008}. This scheme allows for shorter excitation times compared to the resonant excitation schemes using circular polarized pulses \cite{Barth2006b,Nobusada2007,Rasanen2007,Moskalenko_PRA2008}. At $t=0$ we apply  linearly polarized (along the $x$-axis) pulse that creates a time-dependent charge polarization along the $x$-axis (cf. Fig.~\ref{mu_phi_t}). The second pulse is linearly polarized along the $y$-axis and is applied at $t=t_y$. It generates a non-equilibrium current depending on the charge polarization created by the first HCP.  The delay time should be short enough so that  relaxation processes are negligible in between  the pulses. In the IA
The generated non-equilibrium current reads
\begin{equation}\label{Eq:I_dyn}
   I_{\rm neq}=\alpha_y\frac{\mu_x(t_y)}{er_0}I_0\Theta(t-t_y),
\end{equation}
where $\alpha_y$ is the excitation strength of the second HCP and $\mu_x(t_y)$ is the dipole moment created by the first HCP just before the application of the second HCP. Equation \eqref{Eq:I_dyn} delivers
 the total current as well as the individual  currents  in each of the two valleys, in which case
    $\mu_x(t_y)$ should be associated with the charge carriers in the respective valley. Defining the valley current as the difference between the currents flowing in two opposite valleys divided by the particle charge we find for the generated valley current
\begin{equation}\label{Eq:Omega}
   I_{\rm neq}^{\rm v}=\alpha_y\frac{\mu^{+}_x(t_y)-\mu^{-}_x(t_y)}{er_0}\frac{I_0}{e}\Theta(t-t_y).
\end{equation}
On a longer time scale set by the relaxation processes  the non-equilibrium current decays due to dissipation.  Thereby the incoherent electron-phonon scattering  plays usually  the most important role  \cite{Moskalenko_PRB2006,Moskalenko_PRA2008}. Specifically for a free-standing graphene, scattering by flexural phonons is dominant at low temperatures  \cite{Mariani2008}. \\
An example of the dependence of the generated total charge current  on the delay time $t_y$  is depicted in the upper panel of Fig.~\ref{fig2}(b). The oscillating character  of this dependence is determined by the dynamics of the dipole moment generated by the first HCP. The lower panel of Fig.~\ref{fig2}(b) demonstrates the dependence of the generated valley current on the delay time $t_y$. This current arises as a consequence of the different contributions to the total dipole moment from  the two different valleys in presence of a static magnetic flux (here we used $\tilde{\Phi}=1/\pi$) at the time moment $t=t_y$. Comparing the upper and the lower panels we conclude  that tuning the pulses delay  may result in a vanishing total generated current $I_{\rm neq}$  while the generated valley current $I_{\rm neq}^{\rm v}$ is finite.
 It is also possible to create $I_{\rm neq}\neq 0$ with $I_{\rm neq}^{\rm v}=0$. Under the conditions of Fig.~\ref{fig2}(b) the generated currents have the same order of magnitude as the persistent currents. The non-equilibrium contributions are  enhanced however by increasing the HCPs excitation strengths. An increase of the excitation strength $\alpha_x$ of the first HCP beyond  the values around 1 does not lead however
 to an increase of $I_{\rm neq}$ ($I_{\rm neq}^{\rm v}$) under the conditions where $I_{\rm neq}^{\rm v}$ ($I_{\rm neq}$) vanishes because for this, certain delay times are required. In the strong excitation regime the nonlinear oscillations of the dipole moment collapse \cite{Moskalenko_PRB2006} shortly after the excitation (cf. Figs.~\ref{mu_phi_t}(a) and (b) in the range $t/t_0\in[0,2]$). For a further increase of the currents under these conditions $\alpha_y$ should be increased.

The ring charge current is associated with a magnetic dipole moment via $ \vec{M}=\frac{1}{2}\int \vec{r}\times\vec{j}\; {\rm d}^2\vec{r}$, i.e. $M=\pi\int j_\phi r^2 {\rm d}r$. From Eqs.~\eqref{Eq:current_density},\eqref{Eq:R+limit} and \eqref{Eq:R-limit} we infer
for the non-vanishing lowest  order of $W/r_0$
\begin{equation}\label{Eq:magnetic_moment}
    M=\pi r_0^2 I+\pi r_0^2 I_0\sum_{sn}\frac{4s}{(2n+1)^2}Q_{sn},
\end{equation}
where $Q_{sn}=N^+_{sn}-N^-_{sn}$ is the difference in the valley population for fixed $s$ and $n$. For a vanishing total current in the ring and $s=1$, $n=0$,  Eq.~\eqref{Eq:magnetic_moment} simplifies to $M=4Q \pi r_0^2 I_0$. Note, the valley polarized magnetic moment is also a generic feature of the monolayer graphene with a broken inversion symmetry (e.g. due to the action of the substrate potential) \cite{Xiao2007}.
The difference in the valley population in Eq.~\eqref{Eq:magnetic_moment} arises in equilibrium for certain ranges of $\Phi \neq 0$. It can be also generated e.g. by injection of external non-equilibrium carriers to the graphene ring, opening thus a new way  for an ultrafast detection of the valley number. Finally, we note our results
are valid for weak pulses in which case a small angular population around the ground state is created and many-body effects remain subsidiary. Strong excitations go beyond the present model and the influence of many-body interactions may decisively alter the above predictions.
\\
\textit{Conclusion.-} Short linearly polarized asymmetric light pulses trigger a non-equilibrium carrier dynamics in graphene rings threaded by
a magnetic flux. The induced charge polarization  is detectable
by monitoring the emitted radiation.  Delayed pulses with different polarization axes drive   non-equilibrium charge currents and hence an orbital magnetization. For appropriate pulses, equal contributions from both valleys is achievable as well as pure valley currents. The ring magnetization depends on the difference in the valley population.
The predicted effect is operational in presence of tunneling allowing thus for swift injection
 or detection (via ring magnetization) of valley currents in coupled graphene structures, e.g. wires,
 offering new realization of ultrafast valleytronics devices.


\end{document}